\documentclass[11pt,a4paper,pdftex]{article} \usepackage{jcappub}
\pdfoutput=1

\newcommand{\dd}[0]{\ensuremath{{\rm d}}} 						
\newcommand{\Ecut}[0]{\ensuremath{E_{\text{cut}}}}
\newcommand{\tmax}[0]{\ensuremath{t_{\text{max}}}}
\newcommand{\smax}[0]{\ensuremath{s_{\text{max}}}}
\newcommand{\sigmamax}[0]{\ensuremath{\sigma_{\text{max}}}}
\newcommand{\zcr}[0]{\ensuremath{z_{\text{cr}}}}
\newcommand{\zmax}[0]{\ensuremath{z_{\text{max}}}}
\renewcommand{\vec}[1]{\mathbf{#1}}

\usepackage{multicol,multirow}
\usepackage{threeparttable} 									
\usepackage{afterpage} 										

\title{Cosmic ray electrons and positrons from discrete stochastic
  sources}

\author{P. Mertsch}

\affiliation{Rudolf Peierls Centre for Theoretical Physics,\\ 1 Keble
  Road, Oxford OX1 3NP, UK}

\emailAdd{p.mertsch1@physics.ox.ac.uk}

\abstract{The distances that galactic cosmic ray electrons and
  positrons can travel are severely limited by energy losses to at
  most a few kiloparsec, thereby rendering the local spectrum very
  sensitive to the exact distribution of sources in our galactic
  neighbourhood. However, due to our ignorance of the exact source
  distribution, we can only predict the spectrum
  \emph{stochastically}. We argue that even in the case of a large
  number of sources the central limit theorem is not applicable, but
  that the standard deviation for the flux from a random source is
  divergent due to a long power law tail of the probability
  density. Instead, we compute the expectation value and characterise
  the scatter around it by quantiles of the probability density using
  a generalised central limit theorem in a fully analytical way. The
  uncertainty band is asymmetric about the expectation value and can
  become quite large for TeV energies. In particular, the predicted
  local spectrum is marginally consistent with the measurements by
  Fermi-LAT and HESS even without imposing spectral breaks or cut-offs
  at source. We conclude that this uncertainty has to be properly
  accounted for when predicting electron fluxes above a few hundred
  GeV from astrophysical sources.}


\begin{document}
\maketitle

\flushbottom

\section{Introduction}

The interest in the propagation of galactic cosmic ray (GCR) electrons
and positrons from astrophysical sources, like supernova remnants
(SNRs), has recently been revived, mostly in light of contemporary,
partly ``anomalous''
measurements~\cite{Torii:2008xu,Adriani:2008zr,Chang:2008zzr,Collaboration:2008aaa,Abdo:2009zk,Aharonian:2009ah}. In
the context of their possible explanation as exotic signals from dark
matter annihilation or decay, it is usually believed that
astrophysical sources of GCRs only lead to local power law spectra, in
contrast to strong features from exotic sources. This however relies
on a simplified picture of propagation assuming not only a steady
state situation but also a continuous spatial distribution of sources.

On the time and distance scales of GCR propagation, however,
astrophysical sources can be considered \emph{discrete}. Diffusion
renders the fluxes originating from point-like and from extended
sources like old SNRs indistinguishable. Furthermore, although the
details of the injection of particles from SNRs is not fully
understood (for a discussion, see ref.~\cite{Caprioli:2009fv}), the
bulk of the accelerated GCRs is expected to be released with the onset
of the radiative phase, that is within a relatively short time.

For nuclear GCRs, the spectrum depends rather weakly on the exact
distribution of sources in space and time since protons and nuclei
diffuse over distances of kiloparsecs before escaping from the cosmic
ray halo, thereby averaging over the distribution of sources on these
scales. It is therefore admissible to neglect the small scale
distribution and to assume a steady and continuous distribution of
sources. Such a distribution function can be obtained by generalising
information from radio or x-ray surveys of source
candidates~\mbox{\cite{Case:1998qg,Lorimer:2003qc}} and is being used
in many computations of GCR fluxes including the most well-known
numerical \cite{Moskalenko:1997gh,galprop,Evoli:2008dv} and
semi-analytical codes \cite{Maurin:2001sj,Maurin:2002hw}.

The propagation of leptonic GCRs is, however, dramatically different
as electrons and positrons suffer from strong energy losses due to
interactions with the galactic magnetic fields and interstellar
radiation fields (ISRFs). Therefore, the diffusion-loss length,
i.e. the distance electrons and positrons can travel away from the
sources without loosing virtually all energy, is much shorter than for
protons and nuclei. In addition, this distance is energy dependent
such that at high energies, that is above $100 \, \text{GeV}$ or so,
mostly young and nearby sources contribute and the spectrum at these
energies strongly depends on the history and spatial distribution of
sources within a kiloparsec. Therefore, it is \emph{not} admissible to
neglect the nearby small scale distribution.

It has been proposed~\cite{Kobayashi:2003kp} to model the
electron-positron flux by considering a continuous distribution of
sources for distances $\gtrsim 1 \, \text{kpc}$ or ages $\gtrsim 10^5
\, \text{yr}$, and to supplement it by the few known nearby
SNRs. However, this requires that \emph{all} nearby sources are known
from surveys which seems, at least, challenging. First of all, such
studies suffer from various selection effects. Radio surveys, for
instance, are insensitive to SNRs of low surface brightness and also
young but distant ones~\cite{Green:2005yt}. Furthermore, for
individual objects it can be difficult to decide whether they can
contribute to the local electron-positron flux. For example, it has
been suggested that pulsars can accelerate electrons and positrons but
it is not clear if theses high energy particles can escape efficiently
from the surrounding pulsar wind nebulae (PWNe). Recent
observations~\cite{Bamba:2010zk}, however, suggest that this may be
the case. In addition, the prediction of a flux requires accurate
distance information. Considering the uncertainty in the distance
estimates of some of the known SNRs (for example, for RX J1713.7-3946
distance estimates vary between $1$ and $6 \, \text{kpc}$
\cite{Koyama:1997wp,Slane:1999xr}) we cannot expect the distances of
sources yet to be discovered to be much more reliable. Most
importantly however, surveys using electromagnetic radiation will only
tell us about sources on the past light cone. The propagation of
charged GCRs is however dramatically different from the rectilinear
propagation of photons such that much older sources can potentially
contribute. Consequently, relying on surveys we might simply miss old
but nearby sources which could lead to artificial features in the
predicted electron-positron flux (for an illustration of this, see
ref.~\cite{Ahlers:2009ae}).

It thus seems difficult to determine the complete \emph{real}
distribution of sources. However, as we do not expect temporal
variations in the source rate to be too strong, we can generalise the
information about sources on the past light cone to earlier times
thereby building a model of the statistical distribution in both,
position and time. The predictions possible using this statistical
information are however only \emph{stochastic}, that is one can
predict the expectation value for the flux from a statistical ensemble
of sources. For protons and nuclei, the fluctuations around the
expectation value are rather small but for electrons and positrons,
the flux from a particular system in the statistical ensemble, that is
a particular distribution of sources like the one in the Galaxy, will
in general differ from the expectation value. In this sense our
ignorance of the true distribution of sources in age and distance
introduces an uncertainty into the predicted electron-positron flux,
in particular at high energies. This uncertainty can in principle be
investigated by Monte Carlo
methods~\cite{Pohl:1998ug,Strong:2001qp,Swordy:2003ds}, that is by
computing the fluxes from a large number of systems, randomly drawing
the source positions and ages from the statistical model. Here we aim
at a \emph{fully analytical} calculation which will allow us to check
against possible prejudices, e.g. the gaussianity of the probability
density for fluxes. Furthermore, this approach provides the
transparency to trace the propagation of model parameters into the
final result.

The paper is organised as follows. In section~\ref{sec:Propagation} we
briefly review the usual setup for propagation of GCR electrons and
positrons in a purely diffusive model. The calculation for the
expectation value of the flux is outlined in
section~\ref{sec:Expectation} and the determination of the
uncertainties is explained in section~\ref{sec:Uncertainties}. We
present and discuss our results in section~\ref{sec:Discussion} and
conclude in section~\ref{sec:Conclusion}.

\section{Propagation Setup}
\label{sec:Propagation}

The propagation of GCR electrons and positrons is governed by the
diffusive transport equation~\cite{Ginzburg:1990sk},
\begin{equation}
\label{eqn:TransportEquationFore+e-}
\frac{\partial n}{\partial t} - \vec{\nabla} \cdot \left( D \cdot
\vec{\nabla} \right) n - \frac{\partial}{\partial E} \left( b(E) n
\right) = q \, .
\end{equation}
The second and third term on the left hand side describe diffusion,
with the diffusion coefficient $D$, and energy losses, with the energy
loss rate $b(E)$, respectively. We take $b(E) = b_0 E^2$,
parametrising synchrotron radiation and inverse compton scattering in
the Thomson regime. The right hand side denotes injection of electrons
and positrons from the sources. The subdominant secondary production
during the propagation of nuclear GCRs is being ignored. As the
propagation affects positrons in the same way as electrons, the
stochastic effects will be the same for both species, and we consider
only the differential density $n$ of electrons plus
positrons. Furthermore, as the stochasticity effects we are interested
in only show up at higher energies, we neglect convection and
reacceleration which affect the flux below $\sim 10 \,
\text{GeV}$. Finally, solar modulation is expected to modify the
spectrum only marginally at the energies in question, so we neglect
it, too. We consider the transport in a cylinder of radius $\smax$ and
half-height $\zmax$, however only enforcing the boundary condition in
$z$-direction, that is $n(t,s,\zmax) \equiv 0$.

For simplicity, we assume a common spectrum for all
sources\footnote{In fact, the source parameters, e.g. total power
  output, spectral indices and cut-off energies, are in general
  different for different sources. However, this only introduces an
  additional source of uncertainty which we neglect here. In this
  sense, the magnitude of the uncertainty due to the stochasticity of
  source positions and ages is a lower limit on the actual
  uncertainty.}, $Q(E_0)$, such that the source term factorises,
$q(\vec{r},t,E) = \rho(\vec{r},t) Q(E)$. The Green's function with
respect to the source density $\rho(\vec{r},t)$ reads,
\begin{equation}
\mathcal{G}(E,\vec{r},t) = \left( \pi \ell^2 \right)^{-3/2} {\rm
  e}^{-\vec{r}^2/ \ell^2} Q \left( \frac{E}{1 - b_0 E t} \right)
\left( 1 - b_0 E t \right)^{-2} \, ,
\end{equation}
with $\ell = \ell(E,t)$ the diffusion-loss length,
\begin{align}
\ell^2(E, t) = \frac{4 D_0}{b_0 (1 - \delta)} \left[ E^{\delta - 1} -
  \left( \frac{E}{1-b_0 E t} \right)^{\delta - 1} \right] \, .
\end{align}
To satisfy the above boundary condition, we apply the
``mirror-charge'' method of ref.~\cite{Baltz:1998xv}, yielding,
\begin{align}
G (E,\vec{r},t) &= \sum_{n=-\infty}^{\infty} \mathcal{G}(E,\vec{r}_n,
t) \quad \text{where} \quad \vec{r}_n = \left( \begin{array}{c}
  x\\ y\\ (-1)^n z + 2 z_\text{max} n \end{array} \right) \, .
\end{align}
It is useful to factorise the spatial dependence into a dependence on
radius $s$ and $z$,
\begin{align}
G (E,\vec{r},t) &= \sum_{n=-\infty}^{\infty} (\pi \ell^2)^{-3/2}
e^{-\vec{r}_n^2/\ell^2} Q \left( \frac{E}{1 - b_0 E t} \right) (1 -
b_0 E t)^{-2} \\ &= \left( \pi \ell^2 \right)^{-1} e^{-s^2/\ell^2} Q
\left( \frac{E}{1 - b_0 E t} \right) (1 - b_0 E t)^{-2}
\frac{1}{z_\text{cr}} \chi \left( \frac{z}{z_\text{cr}} ,
\frac{\ell^2}{z_\text{cr}^2} \right) \, ,
\label{eqn:Greens}
\end{align}
where the sum has been expressed in terms of the Jacobi theta
function, $\vartheta_3$,
\begin{align}
\chi \left( \hat{z} , \hat{\ell}^2 \right) = \frac{1}{\pi} \left[
  \vartheta_3 \left( \hat{z}, e^{-\hat{\ell}^2} \right) - \vartheta_3
  \left( \hat{z} + \frac{\pi}{2}, e^{-\hat{\ell}^2} \right) \right] \,
,
\end{align}
and $z_\text{cr} = 4 z_\text{max} / \pi$. As most sources as well as
our position are basically in the thin galactic disk, $z \approx 0$
and $\chi( \hat{z} , \hat{\ell}^2 ) \rightarrow \chi( \hat{\ell}^2 )
\equiv \chi( 0 , \hat{\ell}^2 )$.

\section{Expectation Value of the Flux}
\label{sec:Expectation}

The flux from a source that injected a spectrum $Q(E) = Q_0
E^{-\gamma} {\rm e}^{-E/\Ecut}$ of electrons and positrons a time $t$
ago\footnote{In the absence of a reliable time-dependent model of GCR
  escape from a SNR \emph{during} the Sedov Taylor phase we assume
  instantaneous injection. We have checked that for reasonable
  lifetimes $\mathcal{O}(10^4) \, \text{yr}$ this is accurate for all
  except extremely young and nearby sources which are anyway very
  unlikely.} at a distance $s$ from the observer is given by the
Green's function $G$ (cf. eq.~\ref{eqn:Greens}) of the diffusion
equation,
\begin{align}
\label{eqn:GreensDiscreteSrc}
J_i(E) = \frac{c}{4 \pi} G = \frac{c}{4 \pi} \left( \pi \ell^2
\right)^{-1} {\rm e}^{-s^2 / \ell^2} Q_0 E^{-\gamma} (1 - b_0 E
t)^{\gamma - 2} {\rm e}^{-\frac{E}{\Ecut} \frac{1}{1 - b_0 E t} }
\frac{1}{z_\text{cr}} \chi \left( \frac{\ell^2}{z_\text{cr}^2} \right)
\, ,\end{align}
where $c/(4 \pi)$ denotes the ``flux factor'' for relativistic
particles. The flux $J$ of $N$ identical sources at distances $\{ s_i
\}$ and times $\{ t_i \}$ is the sum of the individual fluxes,
\begin{align}
\label{eqn:SumOfFluxes}
\!  J = \sum_{i=1}^N J_i(E) = \frac{c}{4 \pi} \sum_{i=1}^N \left( \pi
\ell^2 \right)^{-1} {\rm e}^{-s_i^2 / \ell^2} Q_0 E^{-\gamma} (1 - b_0
E t_i)^{\gamma - 2} {\rm e}^{-\frac{E}{\Ecut} \frac{1}{1 - b_0 E t} }
\frac{1}{z_\text{cr}} \chi \left( \frac{\ell^2}{z_\text{cr}^2} \right)
.
\end{align}
We note that with the above form of the source spectrum all kinds of
astrophysical sources can be modelled, e.g. SNRs or pulsars.

If the central limit theorem was applicable, at a fixed energy $E$ the
fluxes $J(E)$ for different realisations of $N$ sources drawn from the
same probability density, $f_{s,t}$, would follow a normal
distribution with mean $\mu_J$ and standard deviation $\sigma_J$,
\begin{align}
\mu_J =& \frac{c}{4 \pi} N \mu_G = \frac{c}{4 \pi} N \langle G \rangle
\, , \\ \sigma_J =& \frac{c}{4 \pi} \sqrt{N} \sigma_G = \frac{c}{4
  \pi} \sqrt{N} \sqrt{\langle G^2 \rangle - \langle G \rangle^2} \, ,
\end{align}
were $\langle G^m \rangle$ denotes the moments of the Green's function
$G \equiv G (E, s, t)$,
\begin{align}
\langle G^m \rangle &= \int \dd g \, f_G(g) g^m \, .
\end{align}
As a function of the random variables $s$ and $t$, $G$ itself is a
random variable with the probability density $f_G$. In the case under
consideration, $s$ and $t$ are assumed to be independent random
variables with probability densities $f_s$ and $f_t$, respectively,
and thus the joint probability density $f_{s,t}$ factorises,
$f_{s,t}(s,t) = f_s(s) f_t(t)$.

\begin{figure}[tb]
\begin{tabular*}{\columnwidth}{@{} p{0.49\columnwidth} @{} p{0.02\columnwidth} @{} p{0.49\columnwidth} @{}}
\includegraphics[width=0.45\columnwidth]{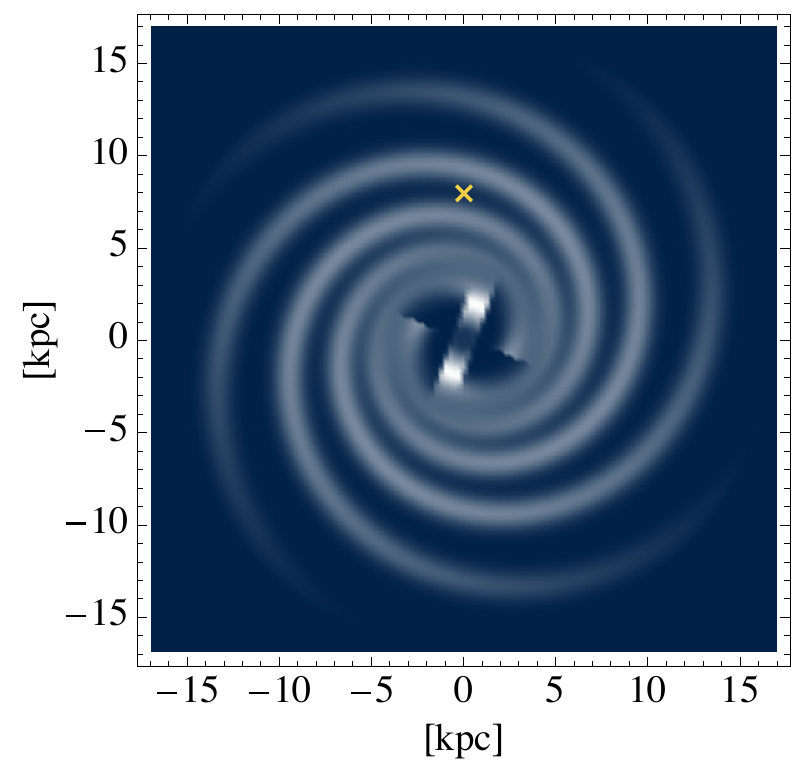} & &
\includegraphics[width=0.45\columnwidth]{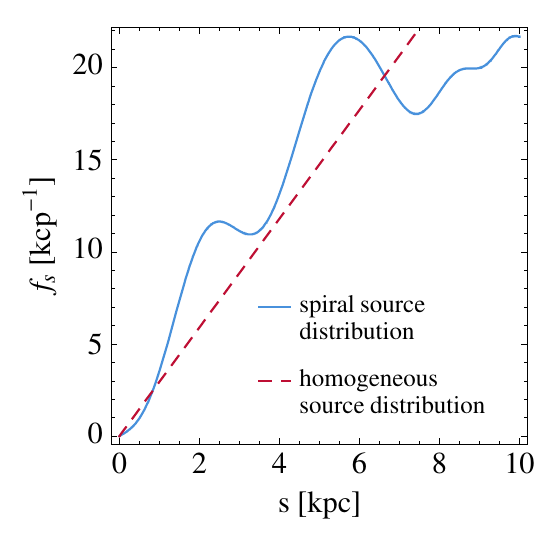} \\
\begin{minipage}[t]{\linewidth}
\caption{The assumed distribution of SNRs in the Galaxy; the cross
  denotes the position of the Sun in between two spiral arms.}
\label{fig:spiral_structure}
\end{minipage}
& &
\begin{minipage}[t]{\linewidth}
\caption{The probability density for the distance of a random SNR from
  the Sun.}
\label{fig:SNR_density}
\end{minipage}
\end{tabular*}
\end{figure}

We assemble a realistic probability density for source distances $s$
by modelling the Galaxy as a logarithmic spiral~\cite{Vallee:2005} and
weighting it in such a way that the distribution in galacto-centric
radius agrees with what is known from radio-surveys of
SNRs~\cite{Case:1998qg}, see figure~\ref{fig:spiral_structure}. The
distance distribution is then obtained by transforming to a
helio-centric coordinate system and averaging over polar angle. For
details, see ref.~\cite{Ahlers:2009ae}. To implement this analytically
complicated expression for $f_s$, we expand it as a power series in
$s^2$, see figure~\ref{fig:SNR_density},
\begin{align}
f_s(s) = \sum_{i=0}^{\infty} a_{2i} s (s^2)^i \, .
\label{eqn:fs}
\end{align}
In practice, we truncate the series after 14 terms and cut off the
source distribution at $\smax$ to prevent the series from
diverging. We emphasise that arbitrary source distributions can be
expanded in such a power series.

Furthermore, we assume that the probability density for the source
ages, e.g. the supernova rate, is constant over the time scales
considered, $\mathcal{O}(100) \, \text{Myr}$ (electrons and positrons
of GeV energies can diffuse over $\mathcal{O}(100) \, \text{Myr}$
before losing their energy but are, of course, subject to escape
losses). Therefore, we take the sources to be equally distributed up
to a maximum time $\tmax = 1/ \left( b_0 E_{\text{min}} \right)$, set
by the minimum energy $E_\text{min}$ to be considered,
\begin{equation}
f_t(t) = \left \{
\begin{array}{cl}
1/\tmax & \text{for } 0 \leq t \leq \tmax \, , \\ 0 & \text{otherwise}
\, .
\end{array}
\right.
\label{eqn:ft}
\end{equation}

Rewriting the $m$-th moment of the Green's function as an integral
over $s$ and $t$, we have,
\begin{align}
\langle G^m \rangle &= \int \dd g \, f_G(g) g^m \\ &=
\int_0^{t_\text{max}} \dd t \int_{0}^{\smax} \dd s f_{s,t}(s,t) \,
G^m(s,t) \label{eqn:DefmthMoment} \\ &= \int_0^{t_\text{max}} \dd t \,
f_{t}(t) \int_{0}^{\smax^2} f_{s}(s) \frac{\dd s^2}{2 s} \, G^m(s,t)
\, .
\end{align}
With eq.~\ref{eqn:fs}, the integral in $s^2$ can be performed and we
substitute $t$ for the injection energy $E_0 = E/(1 - b_0 E t)$. For
$m=1$, this gives,
\begin{align}
\langle G \rangle &= \frac{1}{t_\text{max}} \frac{1}{2 \pi}
\sum_{i=0}^{\infty} a_{2i} \left( - m \right)^{-(1+i)} \int_E^\infty
\frac{\dd E_0}{b_0 E^2} \left( \ell^2 \right)^i \left[ \Gamma \left(
  1+i, \frac{m s^2}{\ell^2} \right) - \Gamma ( 1+i ) \right] \nonumber
\\ & \times Q_0 E_0^{-\gamma} {\rm e}^{-E_0 / \Ecut}
\frac{1}{z_\text{cr}} \chi \left( \frac{\ell^2}{z_\text{cr}^2} \right)
\, .
\end{align}
with $\Gamma$ being the (incomplete) gamma function.

Ignoring the cut-off, $\Ecut \rightarrow \infty$, and expanding the
term $\chi(\hat{\ell}^2)$ for $\ell^2 \ll \zcr^2$ which is only
justified for energies above \mbox{$\sim 50 \, \text{GeV}$} for
$z_\text{max} \simeq 4 \, \text{kpc}$ and amounts to ignoring the
boundary (condition) in the $z$-direction, $(1/\zcr) \chi
(\hat{\ell}^2) \simeq 1 / \sqrt{\pi \hat{\ell}^2}$, we can get
analytical estimates for the moments $\langle G^m \rangle$,
\begin{align}
\langle G^m \rangle &= \frac{1}{t_\text{max}} \frac{1}{2}
\pi^{-\frac{3}{2}m} Q_0^m (b_0 (1-\delta))^{-1} \sum_{i=0}^{\infty}
a_{2i} \Gamma(1+i) m^{-(1+i)} \left( \frac{4 D_0}{b_0 (1 - \delta)}
\right)^{1+i-3m/2} \nonumber \\ & \times E^{-(\delta-1)(\frac{3}{2}m
  -(1+i)) - m \gamma - 1)} \Bigg \{ C_{ik} - \sum_{k=0}^i \frac{(m
  \sigmamax^2)^k}{k!} D_{ik} \Bigg \} \, ,
\end{align}
where
\begin{align}
\! \! C_{ik} &= \frac{\Gamma \left( m\frac{2-\gamma}{\delta-1} -
  \frac{1}{\delta-1} \right) \Gamma \left( 2+i-\frac{3}{2} m
  \right)}{\Gamma \left( 2+i-\frac{3}{2} m +
  m\frac{2-\gamma}{\delta-1} - \frac{1}{\delta-1} \right)} \, , \\ \!
\! D_{ik} &= \Gamma \! \left( m\frac{2-\gamma}{\delta-1} -
\frac{1}{\delta-1} \right) {\rm e}^{-m \sigmamax^2} U \! \left(
m\frac{2-\gamma}{\delta-1} - \frac{1}{\delta-1}, -1-i+k+\frac{3}{2}m,
m \sigmamax^2 \right) \, ,
\end{align}
with $\sigmamax^2 = b_0 (1 - \delta)/(4 D_0) E^{1- \delta} \smax^2$
and Kummer's confluent hypergeometric function $U(a,b,z)$. In
particular, the expectation value of $G$ is,
\begin{align}
\label{eqn:MomentIntegral}
\! \! \langle G \rangle &= \frac{1}{t_\text{max}} \frac{1}{2}
\pi^{-\frac{3}{2}} Q_0 (b_0 (1-\delta))^{-1} \Gamma \left(
\frac{1-\gamma}{\delta-1} \right) \sum_{i=0}^{\infty} a_{2i} \left(
\frac{(4 D_0)}{(b_0 (1 - \delta)} \right)^{-\frac{1}{2}+i}
E^{-(\delta-1)(\frac{1}{2} - i) - \gamma - 1} \nonumber \\ \! \! &
\times \Gamma(1+i) \left\{ \frac{\Gamma \! \left( \frac{1}{2} + i
  \right) }{\Gamma \! \left( \frac{1}{2} + i + \frac{\gamma - 1}{1 -
    \delta} \right) } - {\rm e}^{-\sigmamax^2} \sum_{k=0}^i
\frac{(\sigmamax^2)^k}{k!} { U \left( \frac{1-\gamma}{\delta-1},
  \frac{1}{2}-i+k, \sigmamax^2 \right)} \! \right \} .
\end{align}

For a homogeneous distribution of sources in a disk around the
observer, $a_0 = 2/\smax^2$ and $a_i \equiv 0$ for $i \geq 1$,
\begin{align}
\mu_J = \frac{c}{4 \pi} N \mu_G \simeq \frac{c}{4 \pi}
\frac{1}{\sqrt{4 D_0 b_0 (1 - \delta)}} \frac{N}{\pi \smax^2}
\frac{Q_0}{t_\text{max}} E^{(1 - \delta) / 2 - \gamma - 1}
\frac{\Gamma \left( \frac{1-\gamma}{\delta-1} \right) }{ \Gamma \left(
  \frac{1}{2} + \frac{\gamma - 1}{1 - \delta} \right) } \, .
\end{align}
Choosing, for example, $\delta = 0.6$ and $\gamma = 2.2$ this gives a
spectrum proportional to $E^{-3}$, close to what has been measured by
Fermi-LAT~\cite{Abdo:2009zk}.

For $m=2$, however, the integral in eq.~\ref{eqn:MomentIntegral} and
therefore $\langle G^2 \rangle$ diverge because of the finite
probability density for having an arbitrarily close source which leads
to an arbitrarily large flux. This divergence also occurs for cosmic
ray protons and nuclei as was first pointed out~\cite{Lee:1979zz} in
the context of an energy independent diffusion model.

In principle, this could be cured by introducing a minimum distance
$s_{\text{min}}$ for the distribution $f_s$(s) or a minimum time
$t_{\text{min}}$ for the distribution $f_t(t)$.  In fact, it has been
argued~\cite{Ptuskin:2006zz} that the absence of very young, nearby
sources justifies a lower limit \mbox{$t_{\text{min}}=1/\sqrt{4 \pi
    D(E) \sigma_{\text{SN}}}$} in the time integration
eq.~\ref{eqn:DefmthMoment}, derived from the average distance to the
nearest source assuming a homogeneous source density rate
$\sigma_{\text{SN}} = N / (\pi \smax^2 \tmax)$. This statement can
either be factual, i.e. no flux from such sources has been observed,
or statistical, i.e. the contribution of such sources to the variance
is negligible. As to the former statement, the energies $\gtrsim
\text{few TeV}$ at which such sources would show up are still beyond
the current experimental reach. As to the latter one, the divergence
of the variance exactly shows that such sources contribute
\emph{significantly} in a statistical model.

Without regularisation, the standard deviation of the Green's function
therefore does not exist. In addition, the central limit theorem
cannot be applied anymore and the standard deviation of the
distribution of fluxes does not exist either.  As we will see in the
next section, the reason that the expectation value is finite but not
the variance is that the probability density $f_G(g)$ has a broad
power law tail, $\propto g^{- \alpha - 1}$ with $\alpha \leq 1$. For
such cases, a generalised central limit theorem \cite{Gnedenko:1954}
is applicable: If the probability density $f_G(g)$ behaves like $\left
| g \right |^{-\alpha - 1}$ for $g \rightarrow \infty$, the centred
and normalised sum $X_N$ of $N$ independent and identically
distributed (iid) random variables $G_i$ converges against a stable
distribution $\mathcal{S}(\alpha, \beta, 1, 0,
1)$~\cite{nolan:2010}. In general, the distribution function for
$\mathcal{S}$ is not known analytically but can be calculated as the
inverse Fourier transform of its characteristic function. To determine
the parameters $\alpha$ and $\beta$, we need to find the asymptotic
behaviour of the probability density $f_G(g)$ for large $g$.

\section{Uncertainty of the Flux}
\label{sec:Uncertainties}

\begin{figure}[tb]
\begin{tabular*}{\columnwidth}{@{} p{0.49\columnwidth} @{} p{0.02\columnwidth} @{} p{0.49\columnwidth} @{}}
\begin{center}
\includegraphics[scale=1]{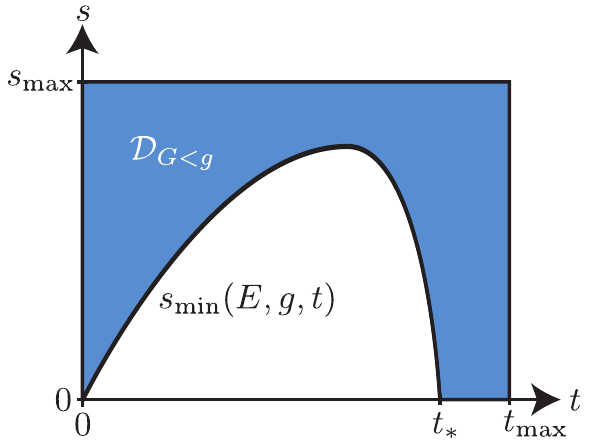}
\end{center}
& &
\begin{center}
\includegraphics[scale=1]{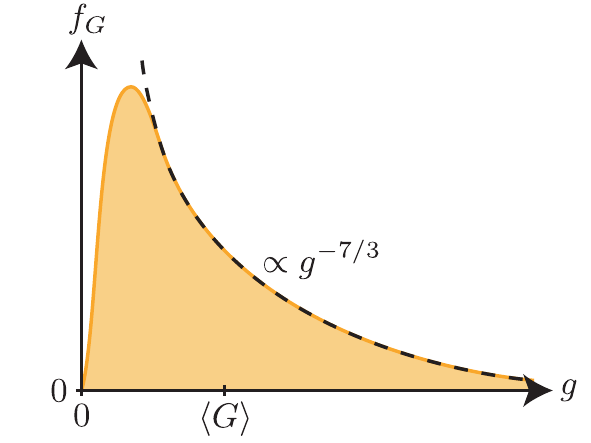}
\end{center}
\\
\begin{minipage}[t]{\linewidth}
\caption{Parameter space for the cumulative distribution function
  $F_G$ for fixed $E$ and $g$. The domain $\mathcal{D}_{G<g}$ is
  limited by $s_{\text{max}}$ from the top and the maximum of $0$ and
  $s_{\text{min}}(E,g,t)$ from the bottom.}
\label{fig:Fig3}
\end{minipage}
& &
\begin{minipage}[t]{\linewidth}
\caption{Probability density $f_G$ for the Green's function. The
  dashed line shows the long power law tail for $g \rightarrow
  \infty$.}
\label{fig:Fig4}
\end{minipage}
\end{tabular*}
\end{figure}

The cumulative distribution function $F_G(g)$, i.e. the probability to
find $G < g$, is given by
\begin{align}
F_G(g) = \iint_{\mathcal{D}_{G < g}} \, \dd t \, \dd s \, f_t(t)
f_s(s) \, ,
\end{align}
where
\begin{align}
\mathcal{D}_{G < g} = \{ (t,s) \, | \, 0 \leq s \leq \smax, 0 \leq t
\leq \tmax, G \leq g \} \, .
\end{align}
With $G \equiv G(E,s,t)$, $G < g$ can be transformed into a condition
on $s$,
\begin{align}
s^2 > s_\text{min}^2 \equiv -\ell^2 \log g + \ell^2 \log \left[ (\pi
  \ell^2)^{-3/2} Q_0 E^{-\gamma} (1 - b_0 E t)^{\gamma - 2} {\rm
    e}^{-E/\left( (1 - b_0 E t) \Ecut \right) } \right] \, ,
\end{align}
see figure~\ref{fig:Fig3}, and hence
\begin{align}
F_G &= \int_0^{t_\text{max}} \, \dd t \, f_t(t) \int_{\max [0;
    s_\text{min}(E,g,t)]}^{\smax} \, \dd s \, f_s(s) \\ &=
\int_0^{t_*} \, \dd t \, f_t(t) \int_{s_\text{min}^2(E,g,t)}^{\smax^2}
\, \frac{\dd s^2}{2 s} \, f_s(s) + \int_{t_*}^{\tmax} \, \dd t \,
f_t(t) \int_0^{\smax^2} \, \frac{\dd s^2}{2 s} \, f_s(s) \, ,
\end{align}
where $s_{\text{min}} \geq 0$ for $0 \leq t \leq t_*$ and
$s_{\text{min}} < 0$ for $t > t_*$. The probability density can then
be obtained by differentiating,
\begin{equation}
f_G = \frac{\dd F_G(g)}{\dd g} = \int_0^{t_*} \dd t \, f_t(t)
\frac{\ell^2}{g} \frac{f_s(s_\text{min})}{2 s_{\text{min}}} =
\frac{1}{2} \frac{1}{\tmax} \frac{1}{g} \sum_{i = 0}^{\infty} a_{2i}
\int_0^{t_*} \dd t \, \ell^2 (s_{\text{min}}^2)^i \, .
\end{equation}

We now substitute,
\begin{align}
\label{eqn:Subst1}
& \lambda^2 = \frac{b_0 (1-\delta)}{4 D_0} E^{1-\delta} \ell^2 = 1 -
(1-b_0 E t)^{1-\delta} \, ,
\end{align}
and find
\begin{align}
f_G &= \frac{1}{2} \frac{1}{\tmax} \frac{1}{g} \sum_{i = 0}^{\infty}
a_{2i} \left( \frac{4 D_0}{b_0 (1-\delta)} \right)^{1+i} (b_0
(1-\delta))^{-1} E^{(\delta-1)(1+i) - 1} \nonumber \\ & \times
\int_0^{\lambda_*^2} \dd \lambda^2 \left( 1 - \lambda^2
\right)^{-\delta/(\delta - 1)} \left( \lambda^2 \right)^{1+i} \\ &
\times \left( - \log g + \log \left[ \left( \frac{4 \pi D_0}{b_0
    (1-\delta)} \right)^{-3/2} Q_0 E^{-\gamma}
  (1-\lambda^2)^{\frac{2-\gamma}{\delta-1}} {\rm e}^{-\frac{E}{\Ecut}
    \left(1 - \lambda_*^2 \right)^{-\frac{1}{1-\delta}} } \right]
-\frac{3}{2} \log \lambda^2 \right)^i \nonumber \, ,
\end{align}
where $\lambda^2_* = 1 - (1-b_0 E t_*)^{1-\delta}$. For $g \rightarrow
\infty$, we anticipate $\lambda^2 \leq \lambda_*^2 \ll 1$ and
therefore neglect the $(1-\lambda^2)$ terms,
\begin{align}
f_G & \simeq \frac{1}{2} \frac{1}{\tmax} \frac{1}{g} \sum_{i =
  0}^{\infty} a_{2i} \left( \frac{4 D_0}{b_0 (1-\delta)} \right)^{1+i}
(b_0 (1-\delta))^{-1} E^{(\delta-1)(1+i) - 1} \nonumber \\ & \times
\int_0^{\lambda_*^2} \dd \lambda^2 \left( \lambda^2 \right)^{1+i}
\left[ - \log g + \log A - \frac{3}{2} \log \lambda^2 \right]^i \, ,
\end{align}
where $A = \left( (4 \pi D_0)/(b_0 (1-\delta) \right)^{-3/2} Q_0
E^{-\gamma} {\rm e}^{-E/\Ecut}$. With the same approximation, we solve
$s_{\text{min}}^2 = 0$ for $\lambda^2_*$,
\begin{equation}
\lambda^2_* \simeq \frac{b_0 (1-\delta)}{4 \pi D_0} E^{1 - \delta -
  \frac{2}{3} \gamma} Q_0^{2/3} g^{-2/3} {\rm e}^{-\frac{2}{3} E/\Ecut
} \, ,
\end{equation}
and we finally compute
\begin{align}
f_G & \simeq \frac{1}{\tmax} \frac{1}{8 D_0} \sum_{i = 0}^{\infty}
a_{2i} \left( \frac{3}{2} \right)^i (2+i)^{-(1+i)} \pi^{-(2+i)} \Gamma
\Big(1+i,(2+i) (\delta-1) \log E \Big) \nonumber \\ & \quad \times
E^{-\delta + (\delta - 1 -\frac{2}{3} \gamma)(2+i)}
Q_0^{\frac{2}{3}(2+i)} {\rm e}^{-\frac{2}{3} (2+i) E/\Ecut }
g^{-\frac{2}{3}(2+i) - 1} \, .
\end{align}

For $g \rightarrow \infty$, the non-zero term with the smallest $i$
determines the asymptotic behaviour, such that for $a_0 \not = 0$, we
have
\begin{equation}
f_G \simeq \frac{1}{\tmax} \frac{1}{8 \pi^2 D_0} \frac{a_0}{2}
E^{-\delta - \frac{4}{3} \gamma} Q_0^{4/3} {\rm e}^{-\frac{4}{3}
  E/\Ecut } g^{-7/3} \, .
\end{equation}
The asymptotic behaviour of the distribution function $F_G$ for $g
\rightarrow \infty$ is therefore
\begin{align}
1 - F_G(g) \sim w g^{-4/3} \quad \text{with} \quad w = \frac{3}{4}
\frac{1}{t_\text{max}} \frac{1}{8 \pi^2 D_0} \frac{a_0}{2} E^{- \delta
  - \frac{4}{3} \gamma} Q_0^{4/3} {\rm e}^{-\frac{4}{3} E/\Ecut } \, ,
\end{align}
and $F_G = 0$ for $g < 0$, see figure~\ref{fig:Fig4}.

The generalised central limit theorem \cite{Uchaikin:1999} then states
that the centred and normalised sum $X_N$,
\begin{align}
X_N = \frac{1}{v_N} \left( \sum_i^N X_i - u_N \right) \, ,
\end{align}
weakly converges to the stable distribution $\mathcal{S}(\alpha,
\beta, 1, 0, 1)$ with $\alpha = 4/3$ and $\beta = 1$. The
normalisation constants are
\begin{align}
\label{eqn:NormalisationConstants}
u_N = N \mu_G \quad \text{and} \quad v_N = \left( \frac{\pi w}{2
  \Gamma(\frac{4}{3}) \sin \left( \frac{\pi}{2} \frac{4}{3} \right) }
\right)^{3/4} N^{3/4} \, .
\end{align}

The expectation value $\mu_J$ for the flux on Earth is therefore still
$c/(4 \pi) N \mu_G$ but instead of the standard deviation, we use
quantiles $x_q$ of the stable distribution to define uncertainty
intervals,
\begin{align}
\mathcal{I}_{68\%} &= \left [ \left( \mu_J + \frac{c}{4 \pi} v_N x_{16
    \%} \right) , \left( \mu_J + \frac{c}{4 \pi} v_N x_{84 \%} \right)
  \right ] \, , \\ \mathcal{I}_{95\%} &= \left [ \left( \mu_J +
  \frac{c}{4 \pi} v_N x_{2.5 \%} \right) , \left( \mu_J + \frac{c}{4
    \pi} v_N x_{97.5 \%} \right) \right ] \, .
\end{align}
For $\mathcal{S}( 4/3, 1, 1, 0, 1)$ the $2.5 \, \%$, $16 \, \%$, $84
\, \%$ and $97.5 \, \%$ quantiles are approximately $-3.6$, $-2.6$,
$1.0$ and $8.7$, respectively. The stable distribution is asymmetric
with respect to its expectation value $x=0$ and so are the uncertainty
bands with respect to $\mu_J$. In particular, the expectation value is
larger than most of the particular realisations of the ensemble and in
this sense not representative which testifies to the long power law
tail of $f_G$. We therefore define a most likely flux $J_{\text{m}}$
through the maximum $x_{\text{max}} \simeq -2.0$ of the stable
distribution,
\begin{equation}
J_{\text{m}} = \frac{c}{4 \pi} \left( u_N + v_N x_{\text{max}} \right)
= \frac{c}{4 \pi} \left( N \mu_G + \left( \frac{\pi w}{2
  \Gamma(\frac{4}{3}) \sin \left( \frac{\pi}{2} \frac{4}{3} \right) }
\right)^{3/4} N^{3/4} x_{\text{max}} \right) \, .
\end{equation}

The energy dependence of both, the quantiles of $J$ and the most
likely flux $J_{\text{m}}$, is contained in $v_N \propto w^{3/4}
\propto E^{-\gamma - 3 \delta / 4} {\rm e}^{-E/\Ecut}$, which is
harder than the expectation value $\mu_J$, so the uncertainty is
growing with energy, as expected. We note the peculiar dependence of
the uncertainty relative to the expectation value, i.e. $v_N/u_N$, on
the source and propagation parameters. In particular, the uncertainty
increases as $Q_0 N^{3/4}$ whereas the expectation goes like $Q_0
N$. The relative fluctuations therefore decrease with the number of
sources as $N^{-1/4}$ which is slower than would have been the case
for the normal central limit theorem where $\sigma_J / \mu_J \propto
N^{-1/2}$.

We note that although the normalisations of the quantiles of $J$ and
of the fluctuations estimated from the square root of the regularised
variance $\langle (\delta N )^2 \rangle^{1/2}$ in
Ref.~\cite{Ptuskin:2006zz} are different, their energy dependence,
$v_N \propto E^{-\gamma - 3 \delta / 4}$, is the same. This is due to
the particular choice for the regulator $t_{\text{min}}$ which
corresponds to a cut-off of the probability density at
$J_{\text{c}}$. It turns out that $J_{\text{c}}$ must itself be a
quantile of the probability density and for sufficiently large
$J_{\text{c}}$ it can be shown that the energy-dependence of $\langle
(\delta N )^2 \rangle^{1/2}$ reduces to $v_N$. Unfortunately, the
definition of $t_{\text{min}}$ does not make manifest which quantile
$\langle (\delta N )^2 \rangle^{1/2}$ corresponds to.

\section{Discussion}
\label{sec:Discussion}

\begin{figure}[p]
\begin{center}
\includegraphics[width=0.7\textwidth]{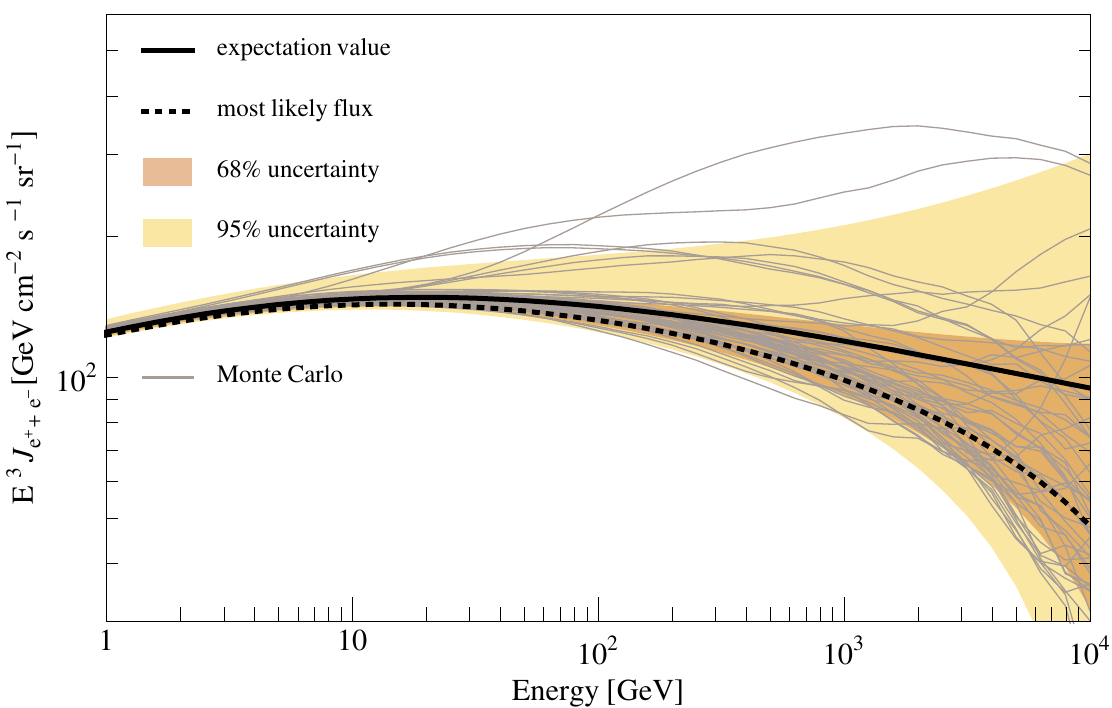}
\end{center}
\caption{Interstellar flux of GCR electrons and positrons from an
  ensemble of source distributions, see eqs.~\ref{eqn:fs}
  and~\ref{eqn:ft}. The solid line denotes the expectation value for
  the sum of fluxes from $N$ discrete, transient sources without
  cut-off, and the dotted line shows the most likely flux,
  corresponding to the maximum of the probability density for
  fluxes. The coloured bands quantify the uncertainty due to our
  ignorance of the real distribution of individual sources, in the
  sense that the fluxes of $68 \, \%$ and $95 \, \%$ of the source
  realisations should lie within the respective bands. The fluxes of
  50 realisations of $N$ individual sources from a Monte Carlo
  calculation are shown by the thin grey lines.}
\label{fig:Spectrum1}
\end{figure}

\begin{figure}[p]
\begin{center}
\includegraphics[width=0.7\textwidth]{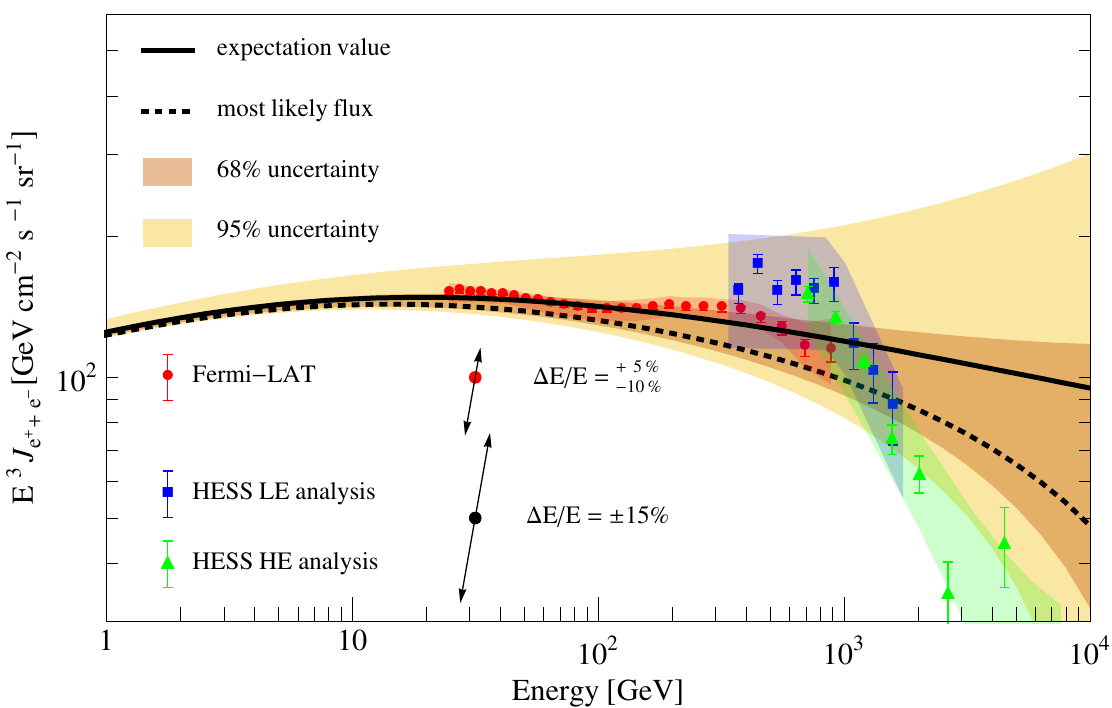}
\end{center}
\caption{Same as figure~\ref{fig:Spectrum1}, but instead of the fluxes
  from the Monte Carlo calculation, we show the electron-positron flux
  as measured by Fermi-LAT~\cite{Abdo:2009zk}, and from the low energy
  (LE)~\cite{Collaboration:2008aaa} and high energy
  (HE)~\cite{Aharonian:2009ah} analyses by HESS. The error bars on the
  data are statistical only, the shaded bands denote the systematic
  uncertainty and the diagonal arrows show the error due to the scale
  uncertainty.}
\label{fig:Spectrum2}
\end{figure}

As an example of the analytical results, we consider the interstellar
electron-positron flux from SNRs distributed according to
eqs.~\ref{eqn:fs} and \ref{eqn:ft}. The parameters used for the source
and diffusion model are summarised in
table~\ref{tbl:Parameters}. Figures~\ref{fig:Spectrum1} and
\ref{fig:Spectrum2} show the calculated expectation value and the
uncertainty bands, defined by the quantiles of the stable
distribution, for $\Ecut \rightarrow \infty$. As explained above, the
spectra are unreliable below $\sim 10 \, \text{GeV}$ as convection and
acceleration have been ignored.

\begin{table}[tb]
\centering
\begin{threeparttable}[b]
\caption{Summary of parameters.}
\label{tbl:Parameters}
\begin{tabular}{c c c}
\hline\hline 
\multicolumn{3}{l}{Diffusion Model}\\
\hline
$D_0$ 			& $10^{28}\,\text{cm}^2\,\text{s}^{-1}$  & \multirow{3}{*}{$\Bigg\}$ \begin{minipage}[c]{0.5\columnwidth} \flushleft from GCR nuclear secondary-to-primary ratios\end{minipage}} \\
$\delta$			& $0.6$ &  \\
$z_\text{max}$				& $3$ 			 $\text{kpc}$ \\
$b_0$ 				& $10^{-16}\,\text{GeV}^{-1} \, \text{s}^{-1}$ &  ISRF and $\vec{B}$ energy densities \\
\hline
\multicolumn{3}{l}{Source Distribution}\\
\hline
$t_\text{max}$		& $3 \times 10^8\,\text{yr}$ 	& from $E_{\text{min}} \simeq 1 \, \text{GeV}$  \\
$N$				&$4.4 \times 10^6$		& from supernova rate, $\smax$ and $t_\text{max}$ \\
\hline
\multicolumn{3}{l}{Source Model}\\
\hline
$Q_0$	 		& $7.8 \times 10^{49}\,\text{GeV}^{-1}$ &\multirow{2}{*}{$\Big\}$ \begin{minipage}[c]{0.5\columnwidth} \flushleft fit to absolute $e^+ + e^-$ flux \end{minipage}} \\
$\gamma$		& $2.2$		& \\
$\smax$			& $10 \, \text{kpc}$		& \\
\hline \hline
\end{tabular}
\end{threeparttable} 
\end{table}

In figure~\ref{fig:Spectrum1} we also show the result of 50 runs of a
Monte Carlo calculation of the electron-positron flux with the same
source and propagation model. Indeed, the fluxes from individual
realisations of the source distribution fluctuate between the
uncertainty bands. We emphasise that the probability density for the
fluxes is not a Gaussian but a more general stable function. We have
checked the statistical interpretation of the uncertainty bands,
i.e. the fluxes of $68 \, \%$ and $95 \, \%$ of the source
realisations lie within the respective bands. As mentioned above, due
to the asymmetry of the probability density $f_G$ around its
expectation value $\mu_G$, most fluxes are smaller than $\mu_G$ but
the ones that are larger deviate more strongly.

\afterpage{\clearpage}

\begin{figure}[p]
\begin{center}
\includegraphics[width=0.7\textwidth]{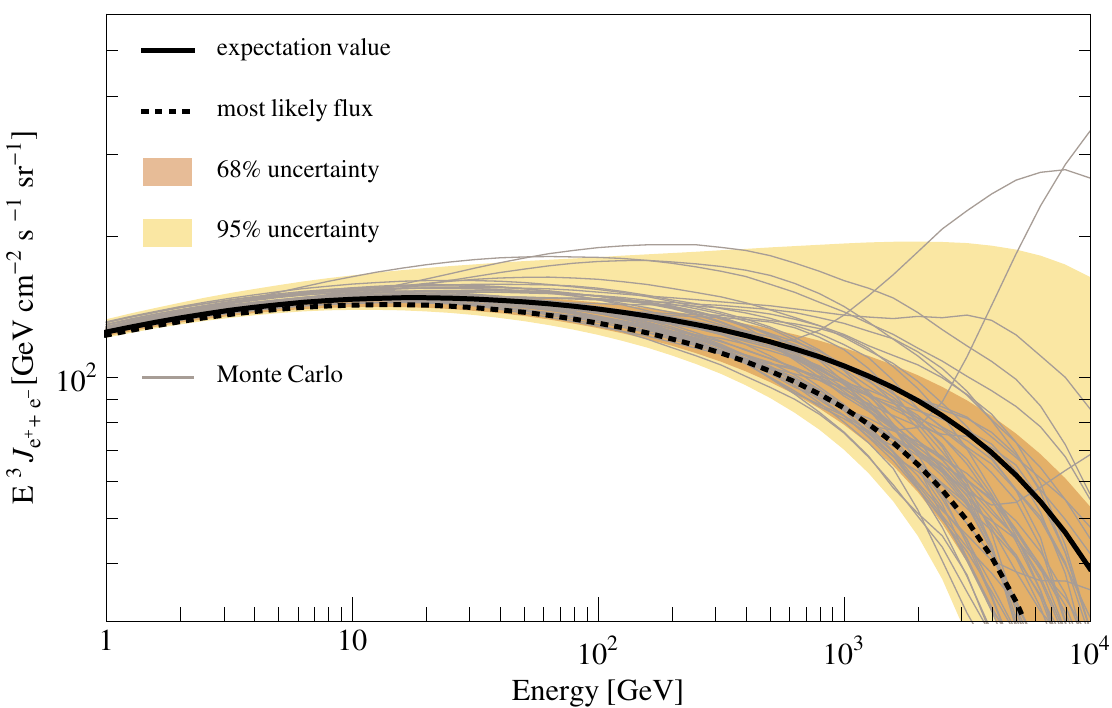}
\end{center}
\caption{Same as figure~\ref{fig:Spectrum1} but assuming a source
  spectrum with cut-off energy $\Ecut = 20 \, \text{TeV}$.}
\label{fig:Spectrum3}
\end{figure}

\begin{figure}[p]
\begin{center}
\includegraphics[width=0.7\textwidth]{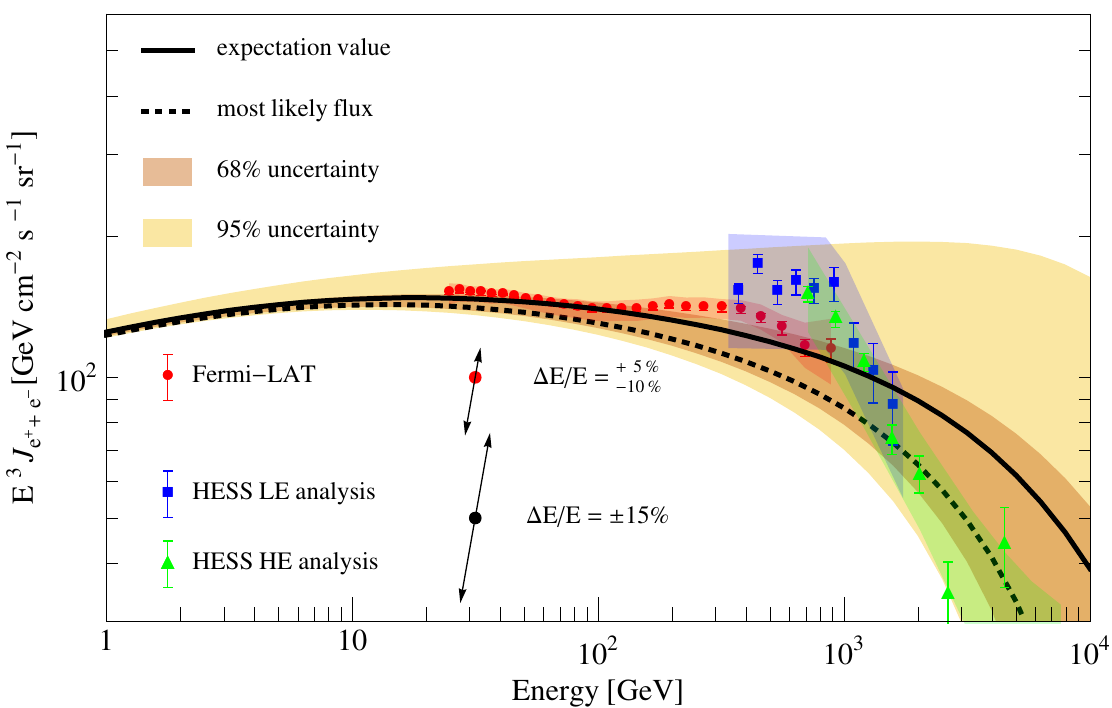}
\end{center}
\caption{Same as figure~\ref{fig:Spectrum2}, but assuming a source
  spectrum with cut-off energy $\Ecut = 20 \, \text{TeV}$.}
\label{fig:Spectrum4}
\end{figure}

What most of the fluxes have in common is a roll-over at a few TeV
compared to the expectation value. This well-known propagation cut-off
is different from a source cut-off and reflects the fact that the
\textit{defacto} youngest source has a finite age
$t_{i_{\text{min}}}$. Demanding $\ell^2 > 0$ then implies a maximum
energy $1 / (b_0 t_{i_{\text{min}}})$, provided that the source $i$ is
close enough to considerably contribute to the total electron-positron
flux. In most cases, a few young sources determine the propagation
cut-off.

In figure~\ref{fig:Spectrum2} we compare our results to the
electron-positron flux as measured by Fermi-LAT~\cite{Abdo:2009zk},
and HESS~\cite{Collaboration:2008aaa,Aharonian:2009ah}. The calculated
expectation value fits the Fermi-LAT data but overshoots the HESS
data, in particular those from the high energy analysis. Choosing a
softer injection, i.e. $\gamma > 2.2$, or imposing a stronger energy
dependence for the diffusion coefficient, i.e. $\delta > 0.6$, would
make the spectrum softer improving the fit to the HESS data but making
the fit to the Fermi-LAT data worse. Within the uncertainty the model
might still reproduce the data, although only the lower edge of the
$95 \, \%$ uncertainty band is marginally consistent with the HESS
data.

The situation can be easily improved on by reinstating the intrinsic
cut-off in the source spectrum, $\Ecut$. For the bulk of (mature)
SNRs, the cut-off energy is naturally expected to lie in the TeV
range~\cite{Reynolds:2008}. Figures~\ref{fig:Spectrum1} and
\ref{fig:Spectrum2} show the calculated expectation value and the
uncertainty bands for $\Ecut = 20 \, \text{TeV}$. The cut-off improves
the fit to the HESS data such that all data lie in the $68 \, \%$
uncertainty band.

\section{Conclusion}
\label{sec:Conclusion}

We have investigated the expectation value in the electron-positron
flux from a statistical ensemble of astrophysical sources as well as
the uncertainty introduced by our ignorance of the distances and ages
of individual sources. We have considered the transport of GCR
electrons and positrons in the usual diffusive setup with the same
power law injection spectrum for all sources. Their average spatial
distribution is modelled combining the distribution of SNRs in
galacto-centric radius with a model of the spiral structure. For this
as for any other source distribution, the expectation value can be
calculated as a sum of power laws in energy where the coefficients are
given by the Taylor coefficients of the source distance
distribution. We found that the probability density $f_G$ of the
Green's function $G$ has a long power law tail. Consequently, the
standard deviation is not defined and the probability density for the
flux is not a Gaussian but a more general stable distribution. We have
quantified the uncertainty around the expectation value by the
quantiles of the stable distribution and have also calculated the
most-likely flux. We find that the level of fluctuations between
different members of the ensemble grows with energy as expected and
that most fluxes exhibit a propagation cut-off at a few TeV. In the
case without source cut-off the uncertainty interval is in agreement
with data from Fermi-LAT and marginally consistent with the
measurements by HESS. This can be improved by invoking a source
cut-off, e.g. at $E_{\text{cut}} = 20 \, \text{TeV}$. The analytical
formulae we have provided allow to consider different types of
astrophysical sources with arbitrary spatial distribution
functions. We emphasise that the uncertainty in the flux is inherent
to the propagation of GCR electrons and positrons and our ignorance of
the positions and ages of individual sources. As such it needs to be
considered when predicting their fluxes from astrophysical sources.

\bibliography{bibliography}{} \bibliographystyle{JHEP}

\end{document}